\RequirePackage{ifpdf}
\documentclass[preprint, preprintnumbrs]{article}
\usepackage{jheppub}
\usepackage{xcolor}
\usepackage{amsmath}
\usepackage{epsfig}
\usepackage{caption}
\usepackage{subcaption}
\usepackage{comment}
\usepackage{hyperref}
\usepackage{soul}
\pdfoutput=1

\usepackage[numbers,sort&compress]{natbib}

\usepackage{upgreek}

\newcommand{\roughly}[1]{\mathrel{\raise.3ex\hbox{$#1$\kern-0.85em
\lower1ex\hbox{$\sim$}}}}

\newcommand{\be}{\begin{equation}}
\newcommand{\bee}{\begin{equation}}
\newcommand{\ee}{\end{equation}}
\newcommand{\beea}{\begin{eqnarray}}
\newcommand{\eea}{\end{eqnarray}}
\newcommand{\bea}{\begin{eqnarray}}

\def\nott#1{\setbox0=\hbox{$#1$}                
   \dimen0=\wd0                                 
   \setbox1=\hbox{/} \dimen1=\wd1               
   \ifdim\dimen0>\dimen1                        
      \rlap{\hbox to \dimen0{\hfil/\hfil}}      
      #1                                        
   \else                                        
      \rlap{\hbox to \dimen1{\hfil$#1$\hfil}}   
      /                                         
   \fi}                                         %

\def\uxsl{\hbox{/\kern-.4000em$u$}}
\def\uxslsm{\hbox{\smaller/\kern-.5600em$u$}}
\def\pxpsl{\hbox{/\kern-.5000em$p$}}
\def\epssl{\hbox{/\kern-.5600em$\epsilon$}}
\def\delsl{\hbox{/\kern-.7000em$\nabla$}}
\def\lxpsl{\hbox{/\kern-.5600em$l$}}
\def\kxpsl{\hbox{/\kern-.5600em$k$}}
\def\qxpsl{\hbox{/\kern-.3900em$q$}}

\def\smath#1{\text{\scalebox{.85}{$#1$}}}
\def\sfrac#1#2{\smath{\frac{#1}{#2}}}

\def\exd{{\rm d}}

\def\UV{{\scriptscriptstyle U\hbox{\kern-0.1em}V}}
\def\PPN{{\scriptscriptstyle P\hbox{\kern-0.1em}P\hbox{\kern-0.1em}N}}
\def\MN{{\scriptscriptstyle M\hbox{\kern-0.1em}N}}
\def\MNP{{\scriptscriptstyle M\hbox{\kern-0.1em}N\hbox{\kern-0.1em}P}}
\def\KK{{\scriptscriptstyle K\hbox{\kern-0.1em}K}}
\def\SM{{\scriptscriptstyle S\hbox{\kern-0.1em}M}}
\def\EH{{\scriptscriptstyle E\hbox{\kern-0.1em}H}}

\def\QCD{{\scriptscriptstyle Q\hbox{\kern-0.1em}C\hbox{\kern-0.1em}D}}
\def\IR{{\scriptscriptstyle I\hbox{\kern-0.1em}R}}
\def\TEV{{\scriptscriptstyle T\hbox{\kern-0.1em}E\hbox{\kern-0.1em}V}}

\def\UV{{\scriptscriptstyle U\hbox{\kern-0.1em}V}}
\def\PPN{{\scriptscriptstyle P\hbox{\kern-0.1em}P\hbox{\kern-0.1em}N}}
\def\MN{{\scriptscriptstyle M\hbox{\kern-0.1em}N}}
\def\MNP{{\scriptscriptstyle M\hbox{\kern-0.1em}N\hbox{\kern-0.1em}P}}
\def\KK{{\scriptscriptstyle K\hbox{\kern-0.1em}K}}
\def\SM{{\scriptscriptstyle S\hbox{\kern-0.1em}M}}
\def\EH{{\scriptscriptstyle E\hbox{\kern-0.1em}H}}

\def\QCD{{\scriptscriptstyle Q\hbox{\kern-0.1em}C\hbox{\kern-0.1em}D}}
\def\IR{{\scriptscriptstyle I\hbox{\kern-0.1em}R}}
\def\TEV{{\scriptscriptstyle T\hbox{\kern-0.1em}E\hbox{\kern-0.1em}V}}
\def\aff{{a\hbox{\kern-0.1em}f\hbox{\kern-0.1em}f}}

\usepackage{wrapfig}

\title{Lindblad evolution as gradient flow}

\author[a,b]{Greg Kaplanek,}
\author[a,c,d]{Alexander Maloney,}
\author[a,b]{Jason Pollack,}
\author[a,c]{and Dylan VanAllen}

\affiliation[a]{Institute for Quantum \& Information Sciences, Syracuse University, NY 13210, USA}

\affiliation[b]{Department of Electrical Engineering and Computer Science, Syracuse University, NY 13210, USA}

\affiliation[c]{Department of Physics, Syracuse University, NY 13210, USA}

\affiliation[d]{Department of Physics, McGill University, Montr\'eal, QC H3A 2T8, Canada}
        
\emailAdd{gkaplane@syr.edu}
\emailAdd{admalone@syr.edu}
\emailAdd{japollac@syr.edu}
\emailAdd{djvanall@syr.edu}

\date{today}

\abstract{We give a simple argument that, for a large class of jump operators, the Lindblad evolution can be written as a gradient flow in the space of density operators acting on a Hilbert space of dimension $D$. We give explicit expressions for the (matrix-valued) eigenvectors and eigenvalues of the Lindblad evolution using this formalism. We argue that in many cases the interpretation of the evolution is simplified by passing from the complex $D^2$-dimensional space of density operators to the real $D^2-1$-dimensional space of Bloch vectors. When jump operators are non-Hermitian the evolution is not in general gradient flow, but we show that it nevertheless resembles gradient flow in two particular ways. Importantly, the steady states of Lindbladian evolution are still determined by the potential in all cases.}

\begin{document}
\maketitle

\section{Introduction}

No system we make measurements on is ever exactly isolated. To the best of our knowledge, the quantum density matrix $\varrho_{\mathrm{world}}$ of everything evolves according to a von Neumann equation of the form
\begin{equation} \label{world}
\frac{\partial \varrho_{\mathrm{world}}}{\partial t}  = - i [ H_{\mathrm{world}}, \varrho_{\mathrm{world}} ] 
\end{equation}
for some Hermitian Hamiltonian $H_{\mathrm{world}}$. In any given measurement, one however only has access to a subset of all available degrees of freedom and so in practice one tracks the quantum state of an open system defined by
\begin{equation}
\rho :=  \underset{\mathrm{env}}{\mathrm{Tr}}\left[ \varrho_{\mathrm{world}} \right]
\end{equation}
where we have taken a partial trace over the unobserved degrees of freedom living in the environment, giving rise to a reduced density matrix $\rho$ of size $D \times D$, and obeying 
\begin{equation} \label{DMcond}
\rho^{\dagger}= \rho \ , \qquad \mathrm{Tr}[\rho]=1 \ , \qquad \rho \geq 0 \ .
\end{equation}
Unlike Eq.~(\ref{world}), the time evolution of $\rho$ can be non-unitary. In practice, when the map taking $\rho(0) \to \rho(t)$ is a quantum channel\footnote{The evolution (\ref{nonMark}) is derived from the Choi \cite{choi1975completely} theorem which ensures that the density matrix is, strictly speaking, {\it completely} positive (conversely any evolution of the form (\ref{nonMark}) can be understood as the partial trace of unitary evolution of the system together with an environment via the Stinespring dilation theorem \cite{stinespring1955positive}).}, one finds the generic form \cite{Lidar:2019qog}
\begin{eqnarray} \label{nonMark}
\frac{\partial \rho}{\partial t} = - i [ H(t) , \rho(t) ] + \sum_{\alpha\beta} c_{\alpha\beta}(t) \Big[ A_{\alpha}(t) \rho(t) A_{\beta}^{\dagger}(t) - \frac{1}{2} \big\{ A_{\beta}^{\dagger}(t) A_{\alpha}(t), \rho(t) \big\} \Big]  
\end{eqnarray}
for a set of time-dependent jump operators $A_{\alpha}(t)$ which are not necessarily Hermitian, a Hermitian system Hamiltonian $H(t)$ and a set of time-dependent real coefficients $\mathfrak{c}(t) = [c_{ij}(t)]$. When the time evolution is a quantum channel, this equation is guaranteed to preserve the conditions (\ref{DMcond}) defining a density matrix, although how these conditions manifest themselves in terms of the structure of $A_{\alpha}(t)$, $H(t)$ and $\mathfrak{c}(t)$ can be highly non-trivial in the general (non-Markovian) case \cite{Breuer:2016zpt,Prudhoe:2022pte}.

Things simplify significantly in the Markovian limit, where the system evolves slowly compared to the timescales of the environment, allowing the dynamics to be treated as memoryless\footnote{Formally, Markovian here means that the superoperator $\mathcal{N}_{t}$ giving $\rho(t) = \mathcal{N}_{t}\big(\rho(0) \big)$ obeys the semi-group property $\mathcal{N}_{t_1+t_2} = \mathcal{N}_{t_1} \circ \mathcal{N}_{t_2}$ and implies that the Lindbladian $\mathcal{L}\big(\rho\big)$ is time-independent so that $\mathcal{N}_{t} = e^{\mathcal{L}t}$.}. In this regime, the general evolution given in Eq.~(\ref{nonMark}) simplifies: the jump operators $A_{\alpha}$ become time-independent constants, and the coefficient matrix $\mathfrak{c} = [c_{\alpha\beta}]$ becomes constant and positive semi-definite.  Diagonalizing the coefficient matrix such that $\mathfrak{c} = U^{\dagger} \gamma U$, one may rotate the jump operators to $B_\alpha := \sqrt{\gamma_\alpha} \sum_{\beta} U_{\beta \alpha} A_{\beta}$ (with $\gamma_\alpha \geq 0$ the positive semi-definite eigenvalues of $\mathfrak{c}$), resulting in the standard form of the Gorini-Kossakowski-Sudarshan-Lindblad (GKSL) \cite{Lindblad:1975ef, Gorini:1976cm} equation, in which the infinitesimal  time evolution of $\rho$ is generated by the Lindbladian $\mathcal{L}$:
\begin{eqnarray} \label{Lindblad_sch}
\frac{\partial \rho}{\partial t} = \mathcal{L}(\rho)\equiv - i [ H , \rho(t) ] + \sum_{\alpha} \Big[ B_{\alpha} \rho B_{\alpha}^{\dagger} - \frac{1}{2} \big\{ B_{\alpha}^{\dagger} B_{\alpha}, \rho(t) \big\} \Big], 
\end{eqnarray}
where there are at most $D^2 - 1$ independent jump operators $B_{\alpha}$ in the second {\it dissipative} term of the Lindblad equation responsible for information loss into the environment. The Lindblad equation turns out to be invariant under unitary reshuffling of the jump operators $B_{\alpha} \to \sum_{\beta} U_{\alpha\beta} B_{\beta}$ as well as the inhomogeneous shift $B_{\alpha} \to B_{\alpha} + z_\alpha \mathbb{I}$ with $H \to H - \frac{i}{2} \sum_{\beta} (z^{\ast}_\beta B_{\beta} - z_\beta B_{\beta}^{\dagger} ) + x \mathbb{I}$ for complex $z_\alpha$ and real $x$. As done in the seminal work \cite{Gorini:1976cm}, one may use this ambiguity in the choice of $H$ and $B_{\alpha}$ to pick the jump operators to be orthonormal with $ \mathrm{Tr}[ B^{\dagger}_{\alpha} B_{\beta} ] = 2 \delta_{\alpha\beta}$ as well as traceless with $\mathrm{Tr}[B_\alpha] =0$. It turns out that choosing traceless jump operators minimizes the size of the dissipative term under several natural super-operator norms, which provides a physical motivation for this choice and selects a canonical form of the Lindblad equation \cite{Hayden:2021fpg}. This minimality property also holds more generally in certain non-Markovian extensions.

Lindblad equations are highly valuable for modelling a wide range of dissipative processes, particularly in fields such as quantum optics, where they describe photon loss, spontaneous emission, and other cavity QED effects. Beyond quantum optics, they have been applied to condensed matter physics, quantum information theory, quantum error correction, holography and cosmology \cite{Burgess:2022nwu}. 

One of the features of Lindblad equations is that they generically take pure states to mixed states (inducing decoherence), via the non-unitary dissipative term in Eq.~(\ref{Lindblad_sch}). To emphasize this point, one may pass to the Dirac (or interaction) picture, so that the unitary terms involving the Hamiltonian are absent. One defines
\begin{eqnarray}
\overline{\rho}(t) := \mathcal{U}^{\dagger}(t) \rho(t) \mathcal{U}(t) \qquad \qquad  \mathrm{and} \qquad \qquad \overline{B}_{\alpha} = \mathcal{U}^{\dagger}(t) B_\alpha \mathcal{U}(t)
\end{eqnarray}
as well as the unitary time evolution operator $\mathcal{U}(t) = \mathcal{T}\exp\left[ - i \int_0^{t} \exd t'\; H(t') \right]$ using the system Hamiltonian $H$ (assuming the evolution starts at time $t=0$), which converts the Lindblad equation to 
\begin{eqnarray} \label{Lindblad}
\frac{\partial \overline{\rho}}{\partial t} = \mathcal{L}(\overline{\rho})\equiv \sum_{\alpha} \Big[ \overline{B}_{\alpha}(t) \overline{\rho}(t) \overline{B}_{\alpha}^{\dagger}(t) - \frac{1}{2} \big\{ \overline{B}_{\alpha}^{\dagger}(t) \overline{B}_{\alpha}(t), \overline{\rho}(t) \big\} \Big] \ . 
\end{eqnarray}
The goal of this work is to develop a simple framework for studying the flow of the above Lindblad equation, centered around the concept of gradient flow\footnote{We work throughout in the Dirac (interaction) picture, which removes the explicit unitary evolution. In this picture, the gradient-flow interpretation focuses on the dissipator structure. Including a Lamb-shift Hamiltonian does not fundamentally change this structure, though in the Schrödinger picture the additional unitary term must be treated explicitly. As emphasized in the context of non-Markovian dynamics \cite{purkayastha2020tunable,Suarez:2024vol}, we also note that Lamb-shift contributions can complicate things, but the essential dissipator structure nevertheless remains amenable to a gradient-flow-like interpretation.}.

\subsection{Relation to prior literature}

In the classical setting, Markov processes are well-known (especially in the context of optimal transport) to admit a gradient flow formulation with respect to classical entropy functions. This perspective has been extended to the quantum setting in recent years, where Lindblad equations --- describing quantum Markovian dynamics --- have likewise been shown to arise as gradient flows of suitable quantum generalizations of entropy \cite{mittnenzweig2017entropic,carlen2017gradient,cao2019gradient,carlen2020non}. Note however that these works rely on a gradient fixed by a non-trivial distance metric, and a scalar entropy function often depending on a known steady-state density matrix $\rho_{\mathrm{eq}}$  and/or dynamics assuming detailed balance. In particular, when the Lindbladian $\mathcal{L}$ (where $\mathcal{L}(\rho)$ is the RHS of Eq.~(\ref{Lindblad_sch})) admits a stationary state $\rho_{\mathrm{eq}}$ ({\it i.e.}\ $\mathcal{L}(\rho_{\mathrm{eq}}) = 0$) and satisfies detailed balance with respect to it, one can define a gradient flow structure for the evolution. The detailed balance condition requires that $\mathcal{L}$ is self-adjoint with respect to the KMS inner product $\langle A,B \rangle_{\mathrm{eq}} := \mathrm{Tr}[A^\dagger B \rho_{\mathrm{eq}}]$ so that, for all operators $A$ and $B$,
\begin{equation}
\langle \mathcal{L}[A], B \rangle_{\mathrm{eq}} = \langle A, \mathcal{L}[B] \rangle_{\mathrm{eq}},
\end{equation}
which ensures the dynamics respects a form of reversibility. Gibbs states, with $\rho_{\mathrm{eq}} \propto e^{-\beta H}$, are a standard example of such equilibrium states. In the simplest cases, detailed balance implies that rates between energy eigenstates obey $\Gamma_{n \to m} / \Gamma_{m \to n} = e^{\beta (E_n - E_m)}$ so there is a notion of reversibility.

The works \cite{mittnenzweig2017entropic,carlen2017gradient,carlen2020non} formalize Lindblad evolution with detailed balance as a gradient flow of the von Neumann (in the case of self-adjoint $\mathcal{L}$) and quantum relative entropy (relative to the steady state $\rho_{\mathrm{eq}}$). In addition, \cite{cao2019gradient} generalizes these constructions to sandwiched R\'enyi divergences assuming GNS-detailed balance. 

In contrast to this body of work, our approach exhibits Lindblad dynamics as a gradient flow with respect to a flat (Euclidean) distance metric, without relying on detailed balance or the previous knowledge of a equilibrium state $\rho_{\mathrm{eq}}$. This simpler structure allows us to frame the dynamics as ordinary gradient descent in operator space, uncoupled from thermodynamic notions of reversibility.

\subsection{Main results}

The main result of the paper is to show that the Lindblad equation (\ref{Lindblad}) can usefully be rewritten as
\begin{equation} \label{Lindblad_PhiR_intro}
\frac{\partial \rho}{\partial t} = \mathcal{L}(\rho)=- \frac{\partial}{\partial \rho^{T}} \Phi(\rho) + R(\rho)
\end{equation}
with the definitions
\begin{eqnarray} 
\Phi(\rho) & := & \frac{1}{4} \sum_{\alpha} \bigg( \big| [B_\alpha, \rho ] \big|^2_{\mathrm{HS}} - \mathrm{Tr}\big( \rho^2  [B_{\alpha},B_{\alpha}^{\dagger}]\big) \bigg) \label{scalar_intro} \\
R(\rho) & := & \frac{1}{2} \sum_{\alpha} \Big( B_{\alpha} \rho B_{\alpha}^{\dagger} - B_{\alpha}^{\dagger} \rho B_{\alpha} \Big),
\end{eqnarray}
where $|O|_{\mathrm{HS}} = \sqrt{ \mathrm{Tr|}[ O^{\dagger} O ]} $ is the Hilbert-Schmidt norm, and where $\frac{\partial}{\partial \rho^{T}} \Phi$ is the matrix gradient of the scalar potential $\Phi$. 

In \S\ref{sec:Herm} we consider the case of Hermitian jump operators with  $B_{\alpha}^{\dagger} = B_{\alpha}$, where the Lindblad equation simplifies to pure gradient flow, as the operator $R(\rho)$ vanishes in this special (degenerate) case, and uniquely determines $\Phi$ (up to constant shifts). This makes the structure of the evolution particularly transparent: when there are enough orthogonal jump operators, the system tends to flow towards the maximally mixed state, which acts as a late-time attractor. This gradient structure also clarifies why the purity of the state always decreases over time in a particularly simple way.

In the more general case where the jump operators are non-Hermitian, $R(\rho)$ does not vanish in general (though it does for qubits), resulting in a more intricate form of evolution. We examine this in \S\ref{sec:nonHerm}, where we show that the right-hand side of Eq.~(\ref{Lindblad_PhiR_intro}) naturally admits a Helmholtz decomposition familiar from vector calculus. In this setting, $R(\rho)$ is a solenoidal matrix field (divergence-free in $\rho$) and crucially, the decomposition into $R$ and $\Phi$ is not unique. Nevertheless, there are two ways to interpret the evolution as a kind of gradient flow. The first, discussed in \S\ref{sec:complex}, involves a complexification of the density matrix. The second, presented in \S\ref{sec:OHHD}, exploits the ambiguity in $R$ and $\Phi$ by choosing a so-called orthogonal Helmholtz-Hodge decomposition such that $ \frac{\partial}{\partial \rho^{T}} \Phi$ is orthogonal to $R$. This condition guarantees that $\Phi$ serves as a Lyapunov function, with the late-time steady states of $\mathcal{L}$ determined solely by the potential $\Phi$.

To make the dynamics more tractable, we also additionally express the Lindblad equation in terms of generalized Bloch vectors for both Hermitian and non-Hermitian jump operators, rewriting the $D \times D$ density matrix $\rho(t)$ in terms of a real $D^2 - 1$ dimensional Bloch vector $\mathbf{a}(t)$. This reformulation is particularly helpful for geometric interpretations and for finding solutions (especially for qudits with $D \geq 3$). To illustrate the physical content of Eq.~(\ref{Lindblad_PhiR_intro}), we explore explicit examples in the Hermitian case in \S\ref{sec:Herm} and in the non-Hermitian case in \S\ref{sec:nonHerm}. 

In what follows we work exclusively in the Dirac (interaction) picture, and so drop the overlines on the operators when the context is clear.

\section{Hermitian jump operators}
\label{sec:Herm}

To begin, we assume that the jump operators appearing in Eq.~(\ref{Lindblad}) are Hermitian\footnote{There is some debate in the literature \cite{alicki2004pure, alicki2007quantum, rivas2012open, carmichael2013statistical, Lidar:2019qog, Correa:2023pwg} about whether Hermitian jump operators can strictly exist in general.}, so that 
\begin{eqnarray} 
B_\alpha^{\dagger} = B_\alpha \ \forall \alpha.
\end{eqnarray}
The main reason we start out with this case is for simplicity. Nevertheless there are many physical situations in which Hermitian jump operators appear in Lindblad equations. For example, in pure dephasing channels the jump operator is Hermitian (a particle number operator), leading to the decay of coherences in the density matrix. Also in the Caldeira-Leggett model of quantum Brownian motion \cite{Caldeira:1981rx}, the jump operator takes the schematic form $B = \sqrt{T} Q + \frac{i}{\sqrt{T}} P$, where $Q$ and $P$ are the system position and momentum operators, with $T$ the temperature. While this operator is generally non-Hermitian, in the high-temperature $T\to \infty$ limit one finds $B \simeq \sqrt{T} Q$ (which {\it is} Hermitian) showing how Hermitian jump can emerge approximately in certain regimes. In this limit, the dynamics is purely dissipative, consistent with the physical picture that the system primarily loses coherence and energy into a hot bath. More generally, whether Hermitian jump operators appear depends on the specific system-environment model and the approximations made in deriving the Lindblad equation. 

\subsection{Gradient flow in matrix space}

In this case, the Lindblad equation (\ref{Lindblad}) in the Dirac picture simplifies to
\begin{eqnarray} \label{Lindblad_H}
\frac{\partial \rho}{\partial t} = - \frac{1}{2} \sum_{i} \big[ B_\alpha, [ B_\alpha, \rho ] \big]  \ .
\end{eqnarray}
Defining the scalar function as in Eq.~(\ref{scalar_intro}) with $B_{\alpha}^{\dagger} = B_{\alpha}$, 
\begin{eqnarray} \label{Vdef_comm}
\Phi(\rho) = \frac{1}{4} \sum_{\alpha} \big| [B_\alpha, \rho ] \big|^2_{\mathrm{HS}},
\end{eqnarray}
one finds generically that $\Phi \geq 0$. The key observation\footnote{This requires some basic matrix calculus where $\frac{\partial }{\partial \rho} f(\rho)$ for a scalar function $f(\rho)$ is short for the matrix with components  $\big[ \frac{\partial }{\partial \rho} f(\rho) \big]_{ij} = \frac{\partial f}{\partial \rho_{ji}}$. With this, one finds $\big[ \frac{ \partial }{\partial \rho^{T}} \rho_{ab} \big]_{ij} = \delta_{ia} \delta_{jb}$ which implies {\it eg.}~$\frac{\partial}{\partial \rho^{T}} \mathrm{Tr}[ X \rho Y \rho ] = X \rho Y + Y \rho X$ and $\frac{\partial}{\partial \rho^{T}} \mathrm{Tr}[ X Y \rho^2 ] = \{ XY,  \rho\}$.} is that the gradient of this scalar function with respect to the reduced density matrix is given by
\begin{equation}
\frac{\partial}{\partial \rho^{T} } \Phi(\rho)  = \frac{1}{2} \sum_{\alpha} \big[ B_{\alpha}(t), [ B_{\alpha}(t), \rho(t) ] \big]
\end{equation}
which means that it simply expresses the Lindblad equation Eq.~(\ref{Lindblad_H}) as a gradient flow equation
\begin{equation} \label{Herm_gradientflow}
\frac{\partial \rho}{\partial t} = \mathcal{L}(\rho) = - \frac{\partial}{\partial \rho^{T}} \Phi(\rho)  \ .
\end{equation}
This formulation is remarkable as it casts the non-unitary evolution of the reduced density matrix in a highly specialized form. One can interpret $\Phi$ as a potential which the time evolution minimizes --- indeed the flow of the state is always in the direction of steepest descent of the potential $\Phi$. This is clearest when studying the {\it purity} of the reduced density matrix, defined as 
\begin{equation} \label{purity}
\gamma := \mathrm{Tr}[\rho^2] \ .
\end{equation}
which is a simple basis-independent scalar which takes the value $1/D \leq \gamma \leq 1$ depending on how mixed $\rho$ is. In the case that $\rho$ is pure (in which case $\rho^2 = \rho$ is a projection) it takes the value 1, and in the maximally mixed case (where $\rho = \mathbb{I}/D$) takes the value $1/D$.

By differentiating the purity, one finds that
\begin{equation} \label{diff_purity}
\frac{\partial \gamma}{\partial t} = 2 \mathrm{Tr}\Big[ \rho \frac{\partial \rho}{\partial t} \Big] =  - 4 \Phi(\rho)  \ .
\end{equation}
which shows in a particularly simple way that purity is always non-increasing over the entire evolution (since $\Phi \geq 0$). The potential determines the instantaneous rate of change of the purity, or equivalently the rate at which the  R\'enyi-2 entropy $S_{2} =- \log(\gamma)$ increases monotonically (similar statements apply to the von Neumann entropy, although the relation is more involved). In the Hermitian case, the potential is therefore the simple quantity that drives entropy production.

\subsubsection{Qubit flows}
\label{sec:qubit_matrix}

As an illustration of the gradient flow picture, let us consider a simplified scenario in which $D=2$ and the jump operators are constant in time. We study the reduced density of a qubit with evolving under the influence of 3 constant Hermitian jump operators $\{ B_{\alpha} \}_{\alpha=1}^3$ (assumed to be distinct).

The gradient flow in Eq.~(\ref{Herm_gradientflow}) suggests that the evolution tends toward minima of the potential. Minimizing the potential means to find the zeros of the right hand side of  Eq.~(\ref{Herm_gradientflow}), and so one seeks the steady states $\rho_{\mathrm{SS}}$ where 
\begin{equation} \label{steady_offlow}
\frac{\partial}{\partial \rho^{T} } \Phi\; \bigg|_{\rho = \rho_{\mathrm{SS}}} = \frac{1}{2} \sum_{\alpha} \big[ B_\alpha, [ B_{\alpha}, \rho_{\mathrm{SS}} ] \big] = 0.
\end{equation}
In the generic case of three linearly independent jump operators, the steady state must therefore be proportional to the identity matrix, and is thus given by the maximally mixed state $\rho_{\mathrm{SS}} = \frac{1}{2}\mathbb{I}$. 

To better understand the dynamics of the solution, we rewrite\footnote{For the qubit case at hand, this is easiest to show by expanding $B_{\alpha}$ in terms of Pauli matrices $\upsigma_{j}$ and the identity $\mathbb{I}$, and then use $[\upsigma_{i},\upsigma_{j}] = 2 i \epsilon_{i j k}$, as well as $\epsilon_{ijk} \epsilon_{mnk} = \delta_{im} \delta_{jn} - \delta_{in} \delta_{jm}$.}
\begin{equation}
- \frac{1}{2} \sum_{\alpha} \big[ B_\alpha, [ B_\alpha, B_{\beta} ] \big] = \ - \sum_{\alpha} \bigg( \mathrm{Tr}[B_{\alpha}^2] B_{\beta} - \mathrm{Tr} [B_{\alpha}B_{\beta}] B_{\alpha} \bigg) \ .
\end{equation}
If one furthermore picks the jump operators  (see the text below Eq.~(\ref{Lindblad_sch})) so they are orthogonal under the Hilbert-Schmidt inner product with $\mathrm{Tr} [B_{\alpha}B_{\beta}]= 2 \delta_{\alpha\beta}$ then one gets
\begin{equation} \label{eigenmatrices}
- \frac{1}{2} \sum_{\alpha} \big[ B_\alpha, [ B_\alpha, B_{\beta} ] \big] = \ - \Gamma_{\beta} B_{\beta}  \qquad \mathrm{with} \qquad \Gamma_{\beta} = \sum_{\alpha \neq \beta} \mathrm{Tr}[B_{\alpha}^2] 
\end{equation}
which shows that $B_{\beta}$ are eigenmatrices of the operator $- \frac{\partial}{\partial \rho^{T} } \Phi(\rho) = - \frac{1}{2} \sum_{\alpha} \big[ B_\alpha, [ B_{\alpha}, \rho ] \big]$ appearing on the RHS of the Lindblad equation with eigenvalues $-\Gamma_{\beta}$. Consequently, the manifestly positive constants $\Gamma_{\beta}$ are the rates with which the solution relaxes to the maximally mixed state.

\subsection{Bloch Vectors}

Quantum mechanics is inherently invariant under changes of bases, and the Lindblad equation of course respects this principle. The Lindblad equation remains unchanged under any unitary transformation $\rho \to U^{\dagger} \rho U$, where $U$ is a fixed element of the unitary group $U(D)$, provided that the operators $B_i$ and $H$ are also rotated accordingly. This suggests that $\rho$ naturally resides within the adjoint representation of $U(D)$.

Given this, it is natural to express $\rho$ in terms of the basis elements of the corresponding Lie algebra. Since $U(D) = U(1) \times SU(D)$ with $SU(D)$ the special unitary group, the Lie algebra $\mathfrak{u}(D)$ is insensitive to the $U(1)$ phase. This means that $\rho$ can be expressed in terms of the Lie algebra $\mathfrak{su}(D)$, which consists of traceless, Hermitian $D \times D$ matrices.

The vector space $\mathfrak{su}(D)$ is $D^{2} - 1$ dimensional, which is consistent with the (real) degrees of freedom contained within a density matrix once the conditions of Hermicity and unit trace are accounted for. Taking $\{ \uplambda_{i} \}_{i=1}^{D^2 - 1}$ to be an orthonormal basis for $\mathfrak{su}(D)$ such that
\begin{equation} \label{lambdaconditions}
\uplambda_j^{\dagger} = \uplambda_j \ , \qquad \mathrm{Tr}\big[ \uplambda_j \big] = 0\ , \qquad \mathrm{Tr}\big[ \uplambda_j \uplambda_k \big] = 2 \delta_{jk} \ ,
\end{equation}
one can then express the density matrix as
\begin{equation} \label{blochN}
\rho = \frac{1}{D} \; \mathbb{I} + \sqrt{ \sfrac{D-1}{2D} } \; a_j \uplambda_{j},
\end{equation}
where we use the Einstein summation convention for repeated indices, and the vector $\mathbf{a} \in \mathbb{R}^{D^2 -1}$ is the {\it generalized Bloch vector} \cite{Arvind:1996rj,jakobczyk2001geometry,byrd2003characterization,kimura2003bloch,kryszewski2006positivity,mendavs2006classification,Goyal:2011xjg} corresponding to the density matrix $\rho$. Given the conditions (\ref{lambdaconditions}), the inversion of Eq.~(\ref{blochN}) is given by
\begin{equation}
a_j = \frac{1}{\sqrt{2 - 2/D}} \; \mathrm{Tr}[ \rho \uplambda_{j} ] \ .
\end{equation}
The Bloch vector is a convenient object to study, as it completely eliminates the redundant degrees of freedom present in the $D \times D$ complex-valued density matrix. Moreover, it enables a direct analysis of the density matrix within the more intuitive Euclidean space, providing a clearer geometric interpretation of the behaviour of the density matrix.

Important to this geometric interpretation is the expression of the purity (\ref{purity}) as
\begin{eqnarray} \label{purityN}
\gamma := \mathrm{Tr}[ \rho^2 ] = \frac{1 + (D-1) |\mathbf{a}|^2}{D} = \begin{cases} \; \frac{1}{D} \qquad & \mathrm{if} \; \mathbf{a} = \mathbf{0} \\
\; \; 1 \qquad & \mathrm{if} \; |\mathbf{a}| = 1 \end{cases}
\end{eqnarray}
This means that $\mathrm{Tr}[\rho^2] \leq 1$ implies that Bloch vectors lie within a hyperball of unit radius $|\mathbf{a}| \leq 1$, with the boundary of the ball corresponding to pure states, and $\mathbf{a} = \mathbf{0}$ corresponding to the maximally mixed state. 

The conditions of Hermiticity and unit trace are built into the Bloch vector definition in Eq.~(\ref{blochN}). The tradeoff is that positivity ($\rho \geq 0$) is generally hard to characterize in terms of the Bloch vector except for the special case of qubits. When $D = 2$, there is a one-to-one correspondence between physical density matrices and the unit ball in $\mathbb{R}^3$, since in that case positivity is equivalent to $\gamma \leq 1$.

For $D \geq 3$, this no longer holds: many vectors $\mathbf{a}$ defined by Eq.~(\ref{blochN}) correspond to non-positive matrices and therefore the set of physical Bloch vectors is a proper subset of the unit ball in $\mathbb{R}^{D^2 - 1}$, and it is non-trivial to describe the structure fo this set \cite{Arvind:1996rj,jakobczyk2001geometry,byrd2003characterization,kimura2003bloch,kryszewski2006positivity,mendavs2006classification}.

Crucial to the study of generalized Bloch vectors is the identity \cite{bossion2021general}
\begin{equation} \label{mult}
\uplambda_j \uplambda_k =  \frac{2}{D} \delta_{jk} \mathbb{I}_D + ( i f_{jk\ell}  + d_{jk\ell} ) \uplambda_\ell,
\end{equation}
where $f_{jk\ell}$ are the (completely antisymmetric) structure constants and $d_{jk\ell}$ are the (completely symmetric) ``$d$-coefficients'' (which turn out to vanish identically for the case $D=2$ and otherwise are non-zero). This means that
\begin{equation} \label{fcomm_def} 
\left[ \uplambda_j , \uplambda_k \right] =  2 i f_{jk\ell} \uplambda_\ell \qquad \mathrm{and} \qquad
\left\{ \uplambda_j , \uplambda_k \right\} = \frac{4}{D} \delta_{jk} \mathbb{I}_D  + 2 d_{jk\ell} \uplambda_\ell \ .
\end{equation}
Given a basis $\{ \uplambda_j \}_{j=1}^{N^2 - 1}$ satisfying (\ref{lambdaconditions}) one can determine $f_{jk\ell}$ and $d_{jk\ell}$ via the relationships
\begin{eqnarray}
f_{jk\ell} =  - \tfrac{i}{4} \mathrm{Tr}\Big( \left[ \uplambda_j , \uplambda_k \right] \lambda_{\ell} \Big) \qquad \mathrm{and} \qquad 
d_{jk\ell} = \tfrac{1}{4} \mathrm{Tr}\Big( \left\{ \uplambda_j , \uplambda_k \right\}  \uplambda_\ell \Big) \ .
\end{eqnarray}
Given these definitions, given two vectors $\mathbf{a},\mathbf{b} \in \mathbb{R}^{D^2 - 1}$ one may define two vector operations \cite{Arvind:1996rj,Bringewatt:2022suo}
\begin{eqnarray}
(\mathbf{a} \wedge \mathbf{b})_{j} = f_{jk \ell} a_{j} b_{k}  \qquad \mathrm{and} \qquad (\mathbf{a} \star \mathbf{b})_{j} = d_{jk \ell} a_{j} b_{k} \ .
\end{eqnarray}
Since these operations depend on a choice of $\{\uplambda_{j}\}_{j=1}^{D^2 - 1}$ they are basis-dependent operations. There is a sense in which the antisymmetric $\wedge$-product is a generalization of the cross product in 3-dimensions, although there isn't a clear analog for the symmetric $\star$-product. Neither of these operations are associative {\it i.e.}\ $( \mathbf{a} \star \mathbf{b} ) \star \mathbf{c} \neq \mathbf{a} \star ( \mathbf{b} \star \mathbf{c} )$, although they are distributive such that $\mathbf{a} \star ( \mathbf{b} + \mathbf{c} ) = \mathbf{a} \star \mathbf{b} + \mathbf{a} \star \mathbf{c}$ and $(\mathbf{a} \star \mathbf{b} )  \cdot \mathbf{c} =  ( \mathbf{b} \star \mathbf{c} ) \cdot \mathbf{a}$ and satisfy for example \cite{Bringewatt:2022suo}
\begin{eqnarray}
(\mathbf{a} \wedge \mathbf{b}) \cdot (\mathbf{c} \wedge \mathbf{d}) =\big( (\mathbf{a} \wedge \mathbf{b} ) \wedge \mathbf{c} \big) \cdot \mathbf{d}
\end{eqnarray}
Since these operations are only sparsely discussed in the literature, we summarize several of their properties in Appendix~\ref{App:WedgeStar}.

\subsection{Gradient flow in the Bloch sphere}

Our goal now is to express the earlier Lindblad equation on the Bloch sphere to facilitate an easier geometric interpretation \cite{ercolessi2001geometry,rossetti2024quantifying}, and to easier find solutions. Assuming the Hermitian jump operators $B^{\alpha}$ are traceless, one can parametrize them in terms of ``jump vectors'' $\mathbf{b}^{(\alpha)}$ such that
\begin{eqnarray}
B_{\alpha} =  \sqrt{ \sfrac{D-1}{2D} } \; b^{(\alpha)}_j \uplambda_{j} \ .
\end{eqnarray}
Using Eq.~(\ref{fcomm_def}) the Lindblad equation (\ref{Lindblad_H}) is trivially found to be\footnote{ This formulation appears in \cite{salgado2004formula} with a variant of this applied to neutron physics in \cite{kerbikov2022neutron}, though at the level of the Bloch vector and without explicit use of gradient flow. }
\begin{eqnarray}
\frac{\partial \mathbf{a}}{\partial t} = \frac{D-1}{D} \; \mathbf{b} \wedge \big( \mathbf{b} \wedge \mathbf{a} \big). 
\end{eqnarray}
It may be equivalently expressed as a gradient flow equation 
\begin{eqnarray}
\frac{\partial \mathbf{a}}{\partial t} = - \frac{D}{D-1} \; \nabla_{\mathbf{a}} \Phi(\mathbf{a}) \qquad \mathrm{with} \qquad \Phi(\mathbf{a}) = \frac{(D-1)^2}{2D^2} \sum_{\alpha} | \mathbf{b}^{(\alpha)} \wedge \mathbf{a} |^2 \ . 
\end{eqnarray}
This complements the earlier gradient flow equation, by mapping it to real Euclidean space. This linear equation can also be written as the matrix equation
\begin{eqnarray} \label{Mmatrix_def}
\frac{\partial \mathbf{a}}{\partial t}  = \mathbb{M}\mathbf{a}(t) \qquad \mathrm{with} \qquad \mathbb{M}_{ij} = \frac{D-1}{D} \sum_{\alpha} f_{i k \ell } f_{\ell m j}  b^{(\alpha)}_{k} b^{(\alpha)}_{m} \ .
\end{eqnarray}
Note furthermore that the matrix $\mathbb{M}$ is closely related the scalar potential $\Phi$ in the sense that it is directly proportional to its Hessian such that $\frac{\partial^2 \Phi}{\partial a_i \partial a_j} = - \frac{D-1}{D} \mathbb{M}_{ij}$.

\subsubsection{Qubit flow in the Bloch sphere}

Let us return to the $D=2$ qubit problem from \S\ref{sec:qubit_matrix} now phrased in terms of Bloch vectors. Here the basis elements of $\mathfrak{su}(2)$ are of course the Pauli matrices, where the structure coefficients are the Levi-Cevita symbol $f_{ijk} = \epsilon_{ijk}$, and the $d$-ceofficients vanish identically $d_{ijk}=0$ (in a sense much of the simplicity that arises for qubits is due to this fact). 

One finds now in this language that the matrix $\mathbb{M}$ from earlier is
\begin{equation} \label{M_qubit1}
\mathbb{M} = \frac{1}{2} \sum_{\alpha} \big( \mathbf{b}^{(\alpha)}\mathbf{b}^{(\alpha)T} - (\mathbf{b}^{(\alpha)} \cdot \mathbf{b}^{(\alpha)} ) \mathbb{I} \big) \ .
\end{equation}
Since $\mathrm{Tr}[B_{\alpha} B_{\beta}] = \frac{1}{2}  \mathbf{b}^{(\alpha)} \cdot  \mathbf{b}^{(\beta)}$ here (since the $d$-coefficients vanish for $\mathfrak{su}(2)$), the orthogonality condition from \S\ref{sec:qubit_matrix} implies that the Bloch vectors are orthogonal with respect to the standard dot product with $\mathbf{b}^{(\alpha)} \cdot  \mathbf{b}^{(\beta)} \propto \delta_{\alpha \beta}$. In this case, the matrix is easily diagonalized, since one now finds
\begin{equation}
\mathbb{M} \mathbf{b}^{(\beta)}  = \ - \Gamma_{\beta} \mathbf{b}^{(\beta)}  \qquad \mathrm{with \ } \Gamma_{\beta} = \tfrac{1}{2} \sum_{\alpha \neq \beta}  \mathbf{b}^{(\alpha)} \cdot \mathbf{b}^{(\alpha)} 
\end{equation}
with the same rates $\Gamma_{\beta}$ as given in Eq.~(\ref{eigenmatrices}), now written in terms of Bloch vectors. The solution (for constant jump operators) is then given by
\begin{equation}
\mathbf{a}(t) =  e^{\mathbb{M} t} \mathbf{a}(0) = \sum_{\beta} e^{- \Gamma_{\beta} t} \frac{\mathbf{b}^{(\beta)} \cdot \mathbf{a}(0) }{ \mathbf{b}^{(\beta)} \cdot \mathbf{b}^{(\beta)}} \; \mathbf{b}^{(\beta)}
\end{equation}
This shows how the solution sinks towards the steady solution at $\mathbf{0}$ at late times $t \to \infty$, corresponding to the maximally mixed state (consistent with the text surrounding Eq.~(\ref{steady_offlow})).
 
In the case that we have a single jump operator the picture slightly changes: consider $B^{(1)} = B$ and $B^{(2)} = B^{(3)} = 0$ so that the earlier matrix (\ref{M_qubit1}) becomes
\begin{eqnarray} 
\mathbb{M} = \frac{1}{2} \big( \mathbf{b} \mathbf{b}^{T} - (\mathbf{b} \cdot \mathbf{b} ) \mathbb{I} \big) 
\end{eqnarray}
with Bloch vector $\mathbf{b}$ corresponding to the traceless $B$. The spectrum of this matrix is different since $\mathbb{M} \mathbf{b} = \mathbf{0}$ and  $\mathbb{M} \mathbf{b}_{\perp} = - \frac{1}{2} ( \mathbf{b} \cdot \mathbf{b} ) \mathbf{b}_{\perp}$ for any vector $\mathbf{b}_{\perp}$ perpendicular to $\mathbf{b}$, where of course the eigenspace corresponding to eigenvalue $ - \frac{1}{2} ( \mathbf{b} \cdot \mathbf{b} ) $ is of dimension 2. The solution in this case is
\begin{eqnarray} \label{singlejump_qubit}
\mathbf{a}(t) & = & \frac{ \mathbf{a}(0) \cdot \mathbf{b} }{ |\mathbf{b}|^2 } \mathbf{b} + \frac{e^{- \frac{1}{2} | \mathbf{b} |^2 t} }{|\mathbf{b}|^2} \;  \big( \mathbf{b} \times \mathbf{a}(0) \big) \times \mathbf{b}
\end{eqnarray}
where $\wedge = \times$ is the usual cross product in $\mathbb{R}^3$. We see that the state sinks towards the projection of the initial state vector $\mathbf{a}(0)$ onto $\mathbf{b}$. 
 
\subsubsection{Qutrit flow on the Bloch sphere}

All of the earlier qubit calculations can be carried out just as easily using either Bloch vectors or matrices. Here, we demonstrate how Bloch vectors can significantly simplify the solution of Lindblad equations in cases where solving them directly in matrix form would be considerably more challenging. Specifically, we consider the case of $D=3$ qutrits with a single jump operator, so that the Lindblad equation takes the form ({\it c.f.} Eq.~(\ref{Mmatrix_def})).
\begin{eqnarray} 
\frac{\partial \mathbf{a}}{\partial t}  = \mathbb{M}\mathbf{a}(t) \qquad \mathrm{with} \qquad \mathbb{M}_{ij} = - \frac{2}{3} \sum_{\alpha} f_{\ell i k } f_{mjk} b^{(\alpha)}_{m} b^{(\alpha)}_{\ell} \ .
\end{eqnarray}
where $\mathbf{a},\mathbf{b} \in \mathbb{R}^{8}$ now. To actually perform calculations in this case it is useful to go an specific basis of $\mathfrak{su}(3)$, where the obvious choice are the well-known Gell-mann matrices\footnote{Although there are other choices: see \cite{harrison20143}.}. In this basis, one has the Cartan-Killing metric tensor
\begin{eqnarray}
f_{ \ell  j k} f_{ m j k } = 3 \delta_{\ell m}\; ,
\end{eqnarray}
which can be used to show for example that $\mathrm{Tr}[\mathbb{M}] = - 2 |\mathbf{b}|^2$. Repeated application of this and other identities found in Appendix \ref{App:WedgeStar} shows that the characteristic equation for the eigenvalues $-\Gamma$ of the matrix $\mathbb{M}$ is given by
\begin{eqnarray}
(-\Gamma)^2 \Big( \; (-\Gamma)^3 + |\mathbf{b}|^2 (-\Gamma)^2 + \tfrac{1}{4} |\mathbf{b}|^4 (-\Gamma) + \tfrac{1}{54} |\mathbf{b}|^6 - \tfrac{1}{18} \big[ (\mathbf{b} \star \mathbf{b}) \cdot \mathbf{b} \big]^2 \; \Big)^2= 0 \ .
\end{eqnarray}
The cubic equation in the above always has real roots (which can be seen by studying the cubic discriminant), and one finds that there are 8 eigenvalues
\begin{eqnarray}
(-\Gamma) \in \{ \; 0, \; 0 , \; r^{(0)}, \; r^{(0)}, \; r^{(1)}, \; r^{(1)}, \; r^{(2)}, \; r^{(2)} \; \} 
\end{eqnarray}
with the cubic roots 
\begin{eqnarray}
r^{(n)} = - \frac{ 2  }{ 3 } |\mathbf{b}|^2 \sin^{2}\bigg( \; \frac{1}{6} \; \mathrm{arccos}\left( \frac{ 6 \; [ \mathbf{b} \star \mathbf{b} \cdot \mathbf{b} ]^2 }{ | \mathbf{b} |^6 } - 1 \right) - \frac{\pi n}{3} \; \bigg) \qquad \mathrm{for} \; n = 0 , 1, 2  \ .
\end{eqnarray}
One can see trivially that the eigenvalues $-\Gamma$ are negative semi-definite, which means that the {\it rates} $\Gamma$ are positive and so the solution sinks towards the steady state. Furthermore, the eigenspace corresponding to the eigenvalue $0$ is spanned by the vectors $\mathbf{b}$ and $\mathbf{b} \star \mathbf{b}$, which means that the steady state is a projection onto this subspace analagous to the qubit case in Eq.~(\ref{singlejump_qubit}). The degeneracy seen in the above eigenvalues is in fact intimately related to the star product: defining a matrix $\mathbb{S}_{\mathbf{b}}$ taking $\mathbb{S}_{\mathbf{b}} \mathbf{a} := \mathbf{b} \star \mathbf{a}$, it turns out that $[\mathbb{M}, \mathbb{S}_{\mathbf{b}}] = 0$. This shows that the single-jump Hermitian operator Lindblad equation is invariant under star products.

\section{Non-Hermitian jump operators}
\label{sec:nonHerm}

In this section we study the full Lindblad equation with non-Hermitian jump operators
\begin{eqnarray}  \label{nonHerm_Lind2}
\frac{\partial \rho}{\partial t} = \mathcal{L}(\rho)=\sum_{\alpha} \Big[ B_{\alpha} \rho B_{\alpha}^{\dagger} - \frac{1}{2} \big\{ B_{\alpha}^{\dagger} B_{\alpha}, \rho \big\} \Big] \ . 
\end{eqnarray}
We see that it does not quite undergo gradient flow in the sense described in \S\ref{sec:Herm}. Nevertheless we show that it exhibits several features which prove useful and similar to gradient flow: in particular, the steady states are still those states which minimize a scalar potential.

\subsection{Lack of gradient flow}

Let us begin by analogy with the earlier Eq.~(\ref{diff_purity}) which expresses the evolution of the purity in terms of a scalar function $\Phi$. We find in the non-Hermitian case that
\begin{equation}
\frac{\partial \gamma}{\partial t} = 2 \mathrm{Tr}\Big[ \rho \frac{\partial \rho}{\partial t} \Big] =  - 4 \Phi(\rho),
\end{equation}
with the definition
\begin{equation} 
\Phi(\rho) := \frac{1}{2} \sum_{\alpha} \bigg( \mathrm{Tr}\big( \rho^2 B_\alpha^{\dagger} B_{\alpha}\big) - \mathrm{Tr}\big( \rho B_{\alpha}^{\dagger} \rho B_{\alpha} \big) \bigg) = \frac{1}{4} \sum_{\alpha} \bigg( \big| [B_\alpha, \rho ] \big|^2_{\mathrm{HS}} - \mathrm{Tr}\big( \rho^2  [B_{\alpha},B_{\alpha}^{\dagger}]\big) \bigg).
\end{equation}
Notice how $\Phi(\rho)$ reduces to Eq.~(\ref{Vdef_comm}) in the Hermitian case where $B_{\alpha}^{\dagger} = B_{\alpha}$. The potential $\Phi(\rho)$ now can have any sign (contrary to the Hermitian case); however, this point will not matter too much, as the potential is in fact no longer unique in this case (for reasons that will become clear below).

Importantly, notice that the gradient of this scalar potential does {\it not} yield the RHS of the Lindblad equation since
\begin{equation}
- \frac{\partial}{\partial \rho^{T}} \Phi(\rho) = \frac{1}{2} \sum_{\alpha} \bigg( B_{\alpha} \rho B^{\dagger}_{\alpha} + B^{\dagger}_{\alpha} \rho B_{\alpha} - \left\{ B_{\alpha}^{\dagger} B_{\alpha}, \rho \right\} \bigg),
\end{equation}
{\it c.f.}~Eq.~(\ref{Herm_gradientflow}). Indeed, the gradient of $\Phi$ generates a string of operators which is necessarily symmetric under the interchange of $B_{\alpha}$ and $B_{\alpha}^{\dagger}$. The anti-symmetric nature of the Lindblad equation under this interchange is why it cannot be expressed as gradient flow. We instead rewrite Eq.~(\ref{nonHerm_Lind2}) as
\begin{equation} \label{Lindblad_PhiR}
\frac{\partial \rho}{\partial t} = \mathcal{L}(\rho)= - \frac{\partial}{\partial \rho^{T}} \Phi(\rho) + R(\rho),
\end{equation}
with the definition
\begin{equation}
R(\rho) := \frac{1}{2} \sum_{\alpha} \Big( B_{\alpha} \rho B_{\alpha}^{\dagger} - B_{\alpha}^{\dagger} \rho B_{\alpha} \Big).
\end{equation}
The second term is anti-symmetric under $B_{\alpha} \to B_{\alpha}^{\dagger}$, which is why it cannot be generated from gradient flow. We can interpret the Lindblad evolution as the combination of a driving force $R(\rho)$ and a conservative force generated by the potential $\Phi$ . 

\subsection{Gradient flow via complexification}
\label{sec:complex}

We can generate the full Lindbladian as a gradient by complexifying $\rho$, {\it i.e.}\ by considering a flow in the larger space of matrices with $\rho$ not necessarily Hermitian. Consider the scalar potential
\begin{equation}
\mathcal{F}(\rho,\rho^{\dagger}) = - \frac{1}{2} \sum_{\alpha} \mathrm{Tr}\big[ B_{\alpha} \rho^\dagger B_{\alpha}^\dagger \rho - B_{\alpha}^\dagger \rho^\dagger B_{\alpha} \rho \big],
\end{equation}
which treats $\rho$ and $\rho^\dagger$ as independent variables.

Differentiating $\mathcal{F}$ with respect to $\rho$ while holding $\rho^\dagger$ fixed yields
\begin{equation}
\frac{\partial \mathcal{F}}{\partial \rho^T} = - \frac{1}{2} \sum_{\alpha} \bigg( B_{\alpha} \rho^\dagger B_{\alpha}^\dagger - B_{\alpha}^\dagger \rho^\dagger B_{\alpha} \bigg). \; 
\end{equation}
If we then proceed to set $\rho^{\dagger} = \rho$, we arrive at
\begin{equation}
- \frac{\partial \mathcal{F}}{\partial \rho^T}\Big|_{\rho=\rho^\dagger} = \frac{1}{2} \sum_{\alpha} \bigg( B_{\alpha} \rho B_{\alpha}^\dagger - B_{\alpha}^\dagger \rho B_{\alpha} \bigg) = R(\rho) \ .
\end{equation}
There is thus a sense in which the RHS of the Lindblad equation is the gradient of $\Phi + \mathcal{F}$, but on the space of \emph{complexified} density matrices. 
Therefore
\begin{equation}
\frac{\partial \rho}{\partial t}= - \frac{\partial }{\partial \rho^T}\Big( \Phi(\rho)+\mathcal{F}(\rho,\rho^{\dagger}) \Big)\; \Big|_{\rho=\rho^\dagger} \ ,
\end{equation}
We note that this equation must be interpreted with some care, because the condition $\rho=\rho^\dagger$ is imposed after taking the derivative $\frac{\partial}{\partial \rho^T}$. 
Thus this is not a  gradient flow in the traditional sense, rather a flow with respect to a gradient in a larger space (the space of complexified density matrices) which is then projected onto the desired subspace (the subspace of positive-semidefinite operators with unit trace) at each step along the flow.

\subsection{Orthogonal Helmholtz-Hodge decomposition}
\label{sec:OHHD}

We first observe that the matrix divergence of the operator $R(\rho)$ in \eqref{Lindblad_PhiR} vanishes:
\begin{equation}
\mathrm{div}_{\mathbf{\rho}} \hspace{0.2mm} R(\rho) := \sum_{ij} \frac{\partial R_{ij}}{\partial \rho_{ij}} = 0 \ .
\end{equation}
This means that $R(\rho)$ is a solenoidal or incompressible matrix field, in the standard sense of multivariable calculus. Hence $R(\rho)$ has no sources or sinks as a function of $\rho$. Thus it is natural to expect that the steady states of the Lindblad equation are determined by the scalar potential alone, with $R$ instead acting analogously to a magnetic field in electromagnetism: it induces rotations but does not by itself create or remove steady states (specifically in the case when $R$ and $\nabla \Phi$ are orthogonal as we soon discuss). While $R(\rho)$ does not contribute simply to purity loss, it may play a role in redistribution processes that could be connected to notions of energy or heat currents \cite{alicki2007quantum} in specific system-bath models. A precise identification would require embedding the Lindblad equation into a microscopic setting, which lies beyond our present scope.

The decomposition of the Lindblad equation into a gradient term and a solenoidal term is not unique: consider any scalar function $\varphi(\rho)$ which satisfies
\begin{equation} \label{LaplacianVarphi}
\Delta_{\rho} \varphi(\rho) := \mathrm{div}_{\mathbf{\rho}} \hspace{0.2mm} \frac{\delta}{\delta \rho^{T}} \varphi(\rho) = \sum_{ij} \frac{\partial^2 \varphi}{\partial^2 \rho_{ij}} = 0 
\end{equation}
{\it i.e.}\ where the operation $\Delta_{\rho}$ is the matrix Laplacian, so that the scalar $\varphi$ is a harmonic function of $\rho$. One can then shift $\Phi$ and the solenoidal field $R$ such that
\begin{equation} \label{gauge1}
\Phi \to \widetilde{\Phi} = \Phi + \varphi \qquad\qquad \mathrm{and} \qquad\qquad R \to \widetilde{R} = R + \sfrac{\delta \varphi}{\delta \rho^{T}}
\end{equation}
where $\widetilde{R}$ is also a solenoidal field satisfying $\mathrm{div}_{\mathbf{\rho}} \hspace{0.2mm} \widetilde{R}(\rho)=0$ because of Eq.~(\ref{LaplacianVarphi}), and the Lindblad equation remains invariant under this transformation, since
\begin{equation} \label{Lindblad_PhiRTILDE}
\frac{\partial \rho}{\partial t} = \mathcal{L}(\rho)= - \frac{\partial}{\partial \rho^{T}} \widetilde{\Phi}(\rho) + \widetilde{R}(\rho) \ .
\end{equation}
{\it cf.}~Eq.~(\ref{Lindblad_PhiR}). Of course, in taking the transformation (\ref{gauge1}) all we've done is added and subtracted $\frac{\delta \varphi}{\delta \rho^{T}}$, so there should be no surprise that the Lindblad equation remains the same. As we describe next, we can exploit this redundancy to obtain a type of generalized gradient flow.

\subsubsection*{Generalizing gradient flow}

We saw above that the Lindblad equation for non-Hermitian jump operators is,  strictly speaking, {\it not} gradient flow. When interpreted correctly as an orthogonal Helmholtz-Hodge decomposition \cite{zhou2012quasi,suda2019construction,suda2020application}, there is a sense in which the Lindblad can nonetheless be interpreted as a generalization of gradient flow. 

To begin, we compute the orbital derivative of $\Phi$ along solutions $\rho(t)$ of the Lindblad equation as
\begin{equation}
\frac{\partial }{\partial t} \Phi\big(\rho(t) \big) = \sum_{jk} \frac{\partial \Phi}{\partial \rho_{jk}} \frac{\partial \rho_{jk}}{\partial t}  = \mathrm{Tr}\bigg( \frac{\partial \Phi}{\partial \rho^{T}} \frac{\exd \rho}{\exd t} \bigg) \ .
\end{equation}
Using the Lindblad equation Eq.~(\ref{Lindblad_PhiR}) one finds that this is equivalently
\begin{equation} \label{orbPhi}
\frac{\partial }{\partial t} \Phi\big(\rho(t) \big) =-  \mathrm{Tr}\bigg( \frac{\partial \Phi}{\partial \rho^{T}} \frac{\partial \Phi}{\partial \rho^{T}} \bigg) +  \mathrm{Tr}\bigg( \frac{\partial \Phi}{\partial \rho^{T}} R \bigg) \ .
\end{equation}
The strategy now is to eliminate the second term in the above equation. As emphasized earlier, the Helmholz-Hodge decomposition is not unique, and one may always shift $\Phi$ and $R$ by a harmonic function $\varphi$ satisfying $\Delta_{\rho} \varphi =0$. This means one can select $\varphi$ so that $\widetilde{\Phi} = \Phi + \varphi$ and $\widetilde{R} = R + \frac{\partial \varphi}{\partial \rho^T}$ give rise to
\begin{equation} \label{orthHH}
\mathrm{Tr}\Big( \frac{\partial \widetilde{\Phi}}{\partial \rho^{T}} \widetilde{R} \Big)  =0, 
\end{equation}
{\it i.e.}~so that the (Hermitian) matrices $\frac{\partial \widetilde{\Phi}}{\partial \rho^{T}}$ and $\widetilde{R}$ are orthogonal under the Hilbert-Schmidt inner product. Note that since the Lindblad equation is linear in $\rho$ it is always true that one can find such a decomposition \cite{suda2020application}. Picking such $\widetilde{\Phi}$ and $\widetilde{R}$ constitutes an {\it orthogonal} Helmholtz-Hodge decomposition, where one now finds that Eq.~(\ref{orbPhi}) gets translated to
\begin{equation}
\frac{\partial }{\partial t} \widetilde{\Phi}\big(\rho(t) \big) =-  \bigg| \frac{\partial \widetilde{\Phi}}{\partial \rho^{T}} \bigg|^2_{\mathrm{HS}} \ .
\end{equation}
This is useful because it shows that solutions $\rho(t)$ ensure that the orbital derivative of $\Phi'$ is non-increasing with $\frac{\exd}{\exd t } \widetilde{\Phi} \leq 0$. This means that the behavior of solutions is governed by the level sets of $\widetilde{\Phi}(\rho(t))$. In particular, we find that the potential is minimized at the equilibrium point, and $\widetilde{\Phi}$ can be interpreted as a Lyapunov function \cite{hahn1963theory}.

Perhaps the most useful property of orthogonal Helmholtz decompositions is that at a fixed point $\rho_{\mathrm{SS}}$ (defined by $\mathcal{L}(\rho_{\mathrm{SS}}) = 0$) one finds that $0 =- \frac{\partial \widetilde{\Phi}}{\partial \rho^{T}} + \widetilde{R} \; |_{\rho_{\mathrm{SS}}}$. It then trivially follows from the orthogonality property (\ref{orthHH}) that
\begin{equation}
\mathcal{L}(\rho_{\mathrm{SS}}) = 0 \qquad \iff \qquad \frac{\partial \widetilde{\Phi}}{\partial \rho^{T}} \Big|_{\rho_{\mathrm{SS}}} = 0
\end{equation}
since the orthogonality property (\ref{orthHH}) means that $| \frac{\partial \widetilde{\Phi}}{\partial \rho^{T}} |^2_{\mathrm{HS}} \hspace{0.3mm} \big|_{\rho_{\mathrm{SS}}} = 0$ at the steady state. Hence we have established that the fixed points of the evolution are determined purely by the potential $\widetilde{\Phi}$.

In practice, however, it appears that it is non-trivial to find potentials satisfying Eq.~(\ref{orthHH}), although there are simple ways of proceeding. For example, a simple harmonic function $\varphi$ that is quadratic in $\rho$ and can be used to shift $(\Phi,R) \to (\widetilde{\Phi},\widetilde{R})$ via Eq.~(\ref{gauge1}) is $\varphi = \mathrm{Tr}[ \rho^2 X ]$ with the constraint $\mathrm{Tr}[X] =0$. This is because $\Delta_{\rho} \mathrm{Tr}[ \rho^2 X ] = 2 D \mathrm{Tr}[X]$ which vanishes when $X$ is traceless and so makes $\varphi$ harmonic in the sense of Eq.~(\ref{LaplacianVarphi}). In principle, one can pick $X$ to so that $(\widetilde{\Phi},\widetilde{R})$ are orthogonal. 

It turns out however that this approach fails to capture the generic case, so we instead reformulate the discussion in terms of Bloch vectors, where the orthogonality conditions are most transparent.

\subsection{Orthogonality in terms of Bloch vectors}

We now pass to the Bloch vector picture, where the somewhat unwieldly matricial calculus gets recast as familiar vector calculus. This is particularly helpful for constructing the conditions required for the potentials $\widetilde{\Phi}$ and $\widetilde{R}$ to be orthogonal in Eq.~(\ref{orthHH}). In the case of non-Hermitian jump operators, the only difference from before is that we must parametrize the ``jump vectors'' $\mathbf{b}^{(\alpha)}$ as complex vectors in $\mathbb{C}^{D^2 - 1}$, such that
\begin{equation}
B_{\alpha}= \sfrac{\mathrm{Tr}[B_{\alpha}]}{D} \; \mathbb{I}_{D} + \sqrt{ \sfrac{D-1}{2D} } \; b^{(\alpha)}_j \uplambda_{j} \quad \qquad \mathrm{or} \qquad \quad B_{\alpha}^{\dagger} = \sfrac{\mathrm{Tr}[B_{\alpha}]^{\ast}}{D} \; \mathbb{I}_{D} + \sqrt{ \sfrac{D-1}{2D} } \; b^{(\alpha)\ast}_j \uplambda_{j}  \ .
\end{equation}
By rewriting the Lindblad equation Eq.~(\ref{nonHerm_Lind2}) as a string of commutators and anti-commutators,
\begin{equation} 
\frac{\partial \rho}{\partial t} = \tfrac{1}{4} \big[  [B_{\alpha}^{\dagger},\rho], B_{\alpha} \big] +  \tfrac{1}{4} \big[  [B_{\alpha},\rho], B_{\alpha}^{\dagger} \big] - \tfrac{1}{4} \big[  \{ B_{\alpha}^{\dagger},\rho\}, B_{\alpha} \big] +  \tfrac{1}{4} \big[  \{ B_{\alpha},\rho\}, B_{\alpha}^{\dagger} \big] \label{better}  \ ,
\end{equation}
one finds that
\begin{small}
\begin{equation} \label{Meq_gennonH}
\frac{\partial \mathbf{a}}{\partial t} = \sum_{\alpha} \Big\{  \sfrac{D-1}{D} \; \mathrm{Re}\Big[ \mathbf{b}^{(\alpha)\ast} \wedge (\mathbf{b}^{(\alpha)} \wedge \mathbf{a} ) \Big] +   \sfrac{D-1}{D} \; \mathrm{Im}\Big[ \mathbf{b}^{(\alpha)\ast} \wedge (\mathbf{b}^{(\alpha)}\star \mathbf{a} ) \Big] + \tfrac{2}{D} \sqrt{ \tfrac{D-1}{2D} } \; \mathrm{Im}\Big[ \mathbf{b}^{(\alpha)\ast} \wedge \mathbf{b}^{(\alpha)} \Big] \Big\}  \ .
\end{equation}
\end{small}\ignorespaces
upon assuming that the jump operators are traceless\footnote{In the case that $\mathrm{Tr}[B_{\alpha}] \neq 0$, one finds an extra term $\tfrac{1}{D} \sqrt{ \tfrac{D-1}{2D} } \; i \;  \Big( \mathrm{Tr}[B_{\alpha}]^{\ast} \; \mathbf{b}^{(\alpha)}- \mathrm{Tr}[B_{\alpha}] \; \mathbf{b}^{(\alpha)\ast} \Big) \wedge \mathbf{a}$. As discussed in the text surrounding Eq.~(\ref{Lindblad_sch}) the canonical choice is to assume tracelessness.}. This reduces to the earlier Eq.~(\ref{Mmatrix_def}) when the jump operator is Hermitian ({\it i.e.}~when $\mathbf{b}^{(\alpha)}$ is real). This can be compactly written as \cite{DiMeglio:2024bir}
\begin{equation} \label{Mmatrix_def}
\frac{\partial \mathbf{a}}{\partial t}  = \mathbb{M}\mathbf{a}(t) +  \mathbf{v} \; ,
\end{equation}
where the $(D^2 - 1) \times (D^2 - 1)$ matrix $\mathbb{M}$ and vector $\mathbf{v} \in \mathbb{R}^{D^2-1}$ have components
\begin{small}
\begin{equation}
\mathbb{M}_{ij} = \tfrac{D-1}{D} \sum_{\alpha} \Big(  f_{i k \ell } f_{\ell m j}  \mathrm{Re}\big[ b^{(\alpha)\ast}_{k} b^{(\alpha)}_{m} \big] +  f_{i k \ell } d_{\ell m j} \mathrm{Im}\big[ b^{(\alpha)\ast}_{k} b^{(\alpha)}_{m} \big] \Big) \ \ \mathrm{and} \ \  v_{i} = \tfrac{2}{D} \sqrt{ \tfrac{D-1}{2D} } \sum_{\alpha} f_{i j k } \mathrm{Im}\big[  b_j^{(\alpha)\ast} b_k^{(\alpha)} \big] \; .
\end{equation}
\end{small}\ignorespaces
Decomposing Eq.~(\ref{Mmatrix_def}) into a gradient and solenoidal term analogous to Eq.~(\ref{Lindblad_PhiRTILDE}) one writes
\begin{equation}
\frac{\partial \mathbf{a}}{\partial t} = - \sfrac{D}{D-1} \nabla_{\mathbf{a}} \widetilde{\Phi}(\mathbf{a}) + \widetilde{\boldsymbol{\mathcal{R}}}(\mathbf{a}),
\end{equation}
where $\widetilde{\Phi}$ is precisely the potential (just expressed in terms of the Bloch vector), and has the general form
\begin{eqnarray} \label{mathcalPR}
\widetilde{\Phi}(\mathbf{a}) := \sfrac{D}{D-1} \left[ \;   \sfrac{1}{2} \mathbf{a}^{T} \mathbb{P} \mathbf{a} + \mathbf{p}^{T} \mathbf{a} \right]
\end{eqnarray}
where $\mathbb{P}$ is an unknown $(D^2-1) \times (D^2 - 1)$ symmetric matrix and $\mathbf{p}$ is an unknown $(D^2 -1)$-dimensional vector. The solenoidal vector function $\widetilde{\boldsymbol{\mathcal{R}}}$ is analogous to $\widetilde{R}$ in Eq.~(\ref{Lindblad_PhiRTILDE}). With this definition one then finds that
\begin{eqnarray} \label{mathcalPR2}
- \sfrac{D}{D-1} \nabla_{\mathbf{a}} \widetilde{\Phi}(\mathbf{a}) = - \mathbb{P} \mathbf{a} - \mathbf{p} \qquad \mathrm{and} \qquad \widetilde{\boldsymbol{\mathcal{R}}}(\mathbf{a}) = ( \mathbb{M} + \mathbb{P}) \mathbf{a} + \mathbf{v} + \mathbf{p}
\end{eqnarray}
in order to ensure that Eq.~(\ref{Mmatrix_def}) continues to hold. Enforcing that $\widetilde{\boldsymbol{\mathcal{R}}}$ is solenoidal with $\nabla_{\mathbf{a}} \cdot \widetilde{\boldsymbol{\mathcal{R}}} = 0$ additionally implies that
\begin{equation} \label{Tr_cond}
\mathrm{Tr}[\mathbb{P}] = - \mathrm{Tr}[\mathbb{M}] = - (D-1) \sum_{\alpha}  \; \mathbf{b}^{(\alpha)\ast} \cdot \mathbf{b}^{(\alpha)} \; ,
\end{equation}
where we used the Cartan-Killing form $f_{njk}f_{mjk} = D \delta_{nm}$ of $\mathfrak{su}(D)$ and the relation $f_{njk}d_{mjk}=0$. The orthogonality condition\footnote{This follows from the original condition Eq.~(\ref{orthHH}), since Hilbert-Schmidt inner products of matrices become inner products of Bloch vectors.} $\nabla_{\mathbf{a}} \widetilde{\Phi} \cdot \widetilde{\boldsymbol{\mathcal{R}}} =0$ then implies
\begin{eqnarray}
\big[ \mathbb{P} \mathbf{a} + \mathbf{p} \big] \cdot \big[ ( \mathbb{M} + \mathbb{P}) \mathbf{a} + \mathbf{v} + \mathbf{p} \big] = 0 \qquad \forall \mathbf{a} \ .
\end{eqnarray}
This translates to the set of equations:
\begin{eqnarray}
\mathbb{M}^{T} \mathbb{P} + \mathbb{P}\mathbb{M} + 2 \mathbb{P}^2 & =& 0 \label{riccati1} \\
\mathbb{M}^{T} \mathbf{p} + \mathbb{P} ( \mathbf{v} + 2\mathbf{p}  )  & = & \mathbf{0} \label{riccati2} \\
\mathbf{p} \cdot (\mathbf{v} + \mathbf{p}) & = & 0 \label{riccati3}
\end{eqnarray}
As pointed out in \cite{suda2020application}, the first equation (for the unknown matrix $\mathbb{P}$) is a special case of the algebraic Riccati equation, which often arises in the context of optimal control theory. In Appendix \ref{App:Riccati}, we outline a general algorithm for developing solutions to Eqs.~(\ref{riccati1})-(\ref{riccati3}) with the constraint $\mathrm{Tr}[\mathbb{P}] = - \mathrm{Tr}[\mathbb{M}]$ from Eq.~(\ref{Tr_cond}), and provide an explicit example in \S\ref{sec:nonH-D}.

\subsection{Examples with non-Hermitian jump operators}

We here provide two explicit examples. First we solve for qubit gradient flow in the case of a non-Hermitian jump operator, and then we consider the orthogonal Helmholtz decomposition for a particular qutrit problem. 

\subsubsection{Qubits with non-Hermitian jump operators}

Let us consider the qubit $D=2$ case again, in the case of a single jump operator which we take to be traceless. Note that qubits are special because the symmetric $d$-coefficients (from Eq.~(\ref{mult})) vanish in this special case. The terms in Eq.~(\ref{Meq_gennonH}) involving $\star$ therefore vanish and $\wedge = \times$ in this case, giving
\begin{eqnarray}
\mathbf{a}'(t) & = & \tfrac{1}{2} \mathrm{Re} \left[ \mathbf{b} \times (\mathbf{b}^{\ast} \times \mathbf{a} ) \right] +  \tfrac{1}{4} \mathbf{b}^{\ast} \times (\mathbf{b} \times \mathbf{a} ) + \tfrac{1}{2} \mathrm{Im}[ \mathbf{b}^{\ast} \times \mathbf{b} ] \ .
\end{eqnarray}
In this simple case, one finds that this is exactly gradient flow, where
\begin{eqnarray}
\mathbf{a}'(t) & = & - \nabla_{\mathbf{a}} \Phi \qquad \mathrm{with}\ \Phi = \Big( \tfrac{1}{4} \big[ \mathbf{b} \times \mathbf{a}(t) \big] \cdot \big[ \mathbf{b}^{\ast} \times \mathbf{a}(t) \big] + \tfrac{1}{2} \mathrm{Im}[ \mathbf{b}^{\ast} \times \mathbf{b} ] \cdot \mathbf{a}  \Big) \ .
\end{eqnarray}
By inspection, one clearly sees that the potential is minimized at a point proportional to $\mathrm{Im}[\mathbf{b}^{\ast} \times \mathbf{b}]$ as we verify more explicitly below. 

The above can be written as a matrix equation
\begin{eqnarray}
\mathbf{a}'(t) = \mathbb{M} \mathbf{a} + \tfrac{1}{2} \mathrm{Im}[\mathbf{b}^{\ast} \times \mathbf{b} ]  \qquad \qquad \mathrm{with} \ \mathbb{M} = \tfrac{1}{4} \mathbf{b}^{\ast} \mathbf{b}^{T}  + \tfrac{1}{4} \mathbf{b} \mathbf{b}^{\dagger} -  \tfrac{1}{2} ( \mathbf{b}^{T} \mathbf{b}^{\ast} ) \mathbb{I}
\end{eqnarray}
Diagonalizing the matrix, one finds eigenvalues $-\Gamma_j$ and eigenvectors $\mathbf{r}_j$ such that
\begin{equation}
  \begin{split}
-\Gamma_{1} & =  - \tfrac{1}{2} \mathbf{b} \cdot \mathbf{b}^{\ast}  \\
-\Gamma_{2} & = - \tfrac{1}{4} \mathbf{b} \cdot \mathbf{b}^{\ast} - \tfrac{1}{4} \sqrt{ (  \mathbf{b} \cdot \mathbf{b} ) ( \mathbf{b}^{\ast} \cdot \mathbf{b}^{\ast} ) } \\
-\Gamma_{3} & = - \tfrac{1}{4} \mathbf{b} \cdot \mathbf{b}^{\ast} + \tfrac{1}{4} \sqrt{ (  \mathbf{b} \cdot \mathbf{b} ) ( \mathbf{b}^{\ast} \cdot \mathbf{b}^{\ast} ) }
  \end{split}
\hspace{25mm}
  \begin{split}
\mathbf{r}_{1} & = \mathbf{b} \times \mathbf{b}^{\ast}  \\
\mathbf{r}_{2}  & = \mathbf{b} - \tfrac{\mathbf{b} \cdot \mathbf{b} }{ \sqrt{ ( \mathbf{b} \cdot \mathbf{b} ) ( \mathbf{b}^{\ast} \cdot \mathbf{b}^{\ast} ) } } \mathbf{b}^{\ast} \\
\mathbf{r}_{3}  & = \mathbf{b} + \tfrac{\mathbf{b} \cdot \mathbf{b} }{ \sqrt{ ( \mathbf{b} \cdot \mathbf{b} ) ( \mathbf{b}^{\ast} \cdot \mathbf{b}^{\ast} ) } } \mathbf{b}^{\ast} \\
  \end{split}
\end{equation}
Notice that in the case of a Hermitian operator, the first two eigenvalues both become $- |\mathbf{b}|^2/2$ and the last eigenvalue becomes 0. Meanwhile the first two eigenvectors formulae stop working as the above formulae predict the eigenvectors are the zero vector, while the last eigenvector is just $\mathbf{b}$. 

The determinant of $\mathbb{M}$ is non-zero and the matrix is invertible, this means that the steady state Bloch vector $\mathbf{a}_{\mathrm{SS}}$ is given by
\begin{eqnarray}
\mathbf{0} = \mathbb{M} \mathbf{a}_{\mathrm{SS}} + \tfrac{1}{2} \mathrm{Im}[ \mathbf{b}^{\ast} \times \mathbf{b}]  \qquad \implies \qquad \mathbf{a}_{\mathrm{SS}} = - \mathbb{M}^{-1} \;  \tfrac{1}{2} \mathrm{Im}[ \mathbf{b}^{\ast} \times \mathbf{b}] =  \frac{ \mathrm{Im}[ \mathbf{b}^{\ast} \times \mathbf{b}] }{\mathbf{b} \cdot \mathbf{b}^{\ast} }
\end{eqnarray}
The length of the steady state (related to the purity via Eq.~(\ref{purityN})) is
\begin{eqnarray}
|\mathbf{a}_{\mathrm{SS}} |^2 = 1 - \frac{ ( \mathbf{b} \cdot \mathbf{b}) ( \mathbf{b}^{\ast} \cdot \mathbf{b}^{\ast} )  }{ ( \mathbf{b} \cdot \mathbf{b}^{\ast})^2 } \ .
\end{eqnarray}
The general solution to $\mathbf{a}'(t) = \mathbb{M} \mathbf{a} + \tfrac{1}{2} \mathrm{Im}[\mathbf{b}^{\ast} \times \mathbf{b} ]$ is given by 
\begin{eqnarray}
\mathbf{a}'(t) & = & \mathbf{a}_{\mathrm{SS}} + e^{ \mathbb{M} t } \big( \mathbf{a}(0) - \mathbf{a}_{\mathrm{SS}} \big) \ = \  \frac{ \mathrm{Im}[\mathbf{b}^{\ast} \times \mathbf{b} ] }{\mathbf{b} \cdot \mathbf{b}^{\ast} } + \sum_{j=1}^3 e^{ - \Gamma_j t } \frac{ \mathbf{r}_j^{\ast} \cdot \mathbf{a}(0)  }{ \mathbf{r}^{\ast}_j \cdot \mathbf{r}_j }  \mathbf{r}_j 
\end{eqnarray}
this shows that the solution relaxes towards the steady state, with timescales $1/\Gamma_j$. It is interesting that one can arbitrarily tune the jump operator $B$ to tune the final steady state --- this has potential application for thermal state preparation.

\subsubsection{Qudits with non-Hermitian jump operators}
\label{sec:nonH-D}

In this subsection, we demonstrate the use of the framework in the calculation of a qudit undergoing Lindblad evolution with the single jump operator $B$ with components $B_{jk} = \gamma \sqrt{j} \delta_{j,j+1}$, where $\gamma > 0$ is a constants with units of energy. This corresponds to amplitude damping \cite{Otten:2020xmo}, where the jump operator is the annihilation operator for a $D$-level system. Using the standard Gell-mann matrices one finds for example with $D=3$ that
\begin{equation}
B = \gamma \left[\begin{matrix} 0 & 1 & 0 \\
 0 & 0 & \sqrt{2} \\
  0 & 0 & 0 \end{matrix} \right] \qquad \implies \qquad \mathbf{b} = \frac{\sqrt{3}}{2} \gamma \; \left( 1, i, 0, 0, 0, \sqrt{2}, \sqrt{2} i, 0  \right) \ .
\end{equation}
The Lindblad equation for this choice of jump operator can be written in terms of Bloch vectors as the equation
\begin{equation}
\frac{\exd \mathbf{a}}{\exd t} = \mathbb{M} \mathbf{a}(t) + \mathbf{v} \quad \mathrm{with} \ \mathbb{M} = \left[
\begin{smallmatrix}
 -\gamma ^2/2 & 0 & 0 & 0 & 0 & \sqrt{2} \gamma ^2 & 0 & 0 \\
 0 & -\gamma ^2/2 & 0 & 0 & 0 & 0 & \sqrt{2} \gamma ^2 & 0 \\
 0 & 0 & -\gamma ^2 & 0 & 0 & 0 & 0 & \sqrt{3} \gamma ^2 \\
 0 & 0 & 0 & -\gamma ^2 & 0 & 0 & 0 & 0 \\
 0 & 0 & 0 & 0 & -\gamma ^2 & 0 & 0 & 0 \\
 0 & 0 & 0 & 0 & 0 & - 3 \gamma ^2/2 & 0 & 0 \\
 0 & 0 & 0 & 0 & 0 & 0 & - 3 \gamma ^2/2 & 0 \\
 0 & 0 & 0 & 0 & 0 & 0 & 0 & -2 \gamma ^2 \\
\end{smallmatrix} \right] \ \mathrm{and} \  \mathbf{v} = \left[\begin{smallmatrix} 0 \\ 0 \\ 0 \\ 0 \\ 0 \\ 0 \\ 0 \\ \gamma ^2 \end{smallmatrix}  \right] \ .
\end{equation}
Note that $\det \mathbb{M} = \frac{9}{8} \gamma^{16} \neq 0 $ which means the matrix $\mathbb{M}$ is invertible: the steady-state is clearly given by
\begin{equation} \label{dampSS}
\mathbf{a}_{\mathrm{SS}} = - \mathbb{M}^{-1} \mathbf{v} =  \left[\begin{smallmatrix} 0 \\ 0 \\ {\sqrt{3}}/{2} \\ 0 \\ 0 \\ 0 \\ 0\\ 1/2 \end{smallmatrix}  \right] 
\end{equation}
which using Eq.~(\ref{blochN}) means that the steady-state is given by the density matrix $\rho_{\mathrm{SS}} = \frac{1}{3} \mathbb{I} + \tfrac{1}{2} \uplambda_{3} + \tfrac{1}{2\sqrt{3}} \uplambda_{8}=\mathrm{diag}(1,0,0)$  corresponding to a pure state, in which the system ends up in the ground state of the computational basis.

Note however, that one may use instead the orthogonal Helmholtz-Hodge decomposition for studying the behavior of the solution. Indeed one finds that 
\begin{equation}
\mathbb{P} = \left[
\begin{smallmatrix}
 {\gamma ^2}/{3} & 0 & 0 & 0 & 0 & -{\gamma ^2}/{3 \sqrt{2}} & 0 & 0 \\
 0 & {\gamma ^2}/{3} & 0 & 0 & 0 & 0 & -{\gamma ^2}/{3 \sqrt{2}} & 0 \\
 0 & 0 & {3 \gamma ^2}/{4} & 0 & 0 & 0 & 0 & -{ \sqrt{3} \gamma ^2 }/{4}  \\
 0 & 0 & 0 & \gamma ^2 & 0 & 0 & 0 & 0 \\
 0 & 0 & 0 & 0 & \gamma ^2 & 0 & 0 & 0 \\
 - {\gamma ^2}/{3 \sqrt{2}} & 0 & 0 & 0 & 0 & {5 \gamma ^2}/{3} & 0 & 0 \\
 0 & -{\gamma ^2}/{3 \sqrt{2}} & 0 & 0 & 0 & 0 & {5 \gamma ^2}/{3} & 0 \\
 0 & 0 & - {\sqrt{3} \gamma ^2}/{4}  & 0 & 0 & 0 & 0 & {9 \gamma ^2}/{4} 
\end{smallmatrix}
\right] \qquad \mathrm{and} \qquad \mathbf{p} = \left[ \begin{smallmatrix} 0 \\ 0\\ -{\sqrt{3} \gamma ^2}/{4} \\ 0 \\ 0 \\ 0 \\ 0 \\ - {3 \gamma ^2}/{4}  \end{smallmatrix} \right]
\end{equation} 
solves Eqs.~(\ref{riccati1})-(\ref{riccati3}) for this particular choice of jump operator. Using Eq.~(\ref{mathcalPR}) this means that the potential is
\begin{eqnarray}
\widetilde{\Phi}(\mathbf{a}) & := & \sfrac{3}{2} \left[ \;   \sfrac{1}{2} \mathbf{a}^{T} \mathbb{P} \mathbf{a} + \mathbf{p}^{T} \mathbf{a} \; \right] \\
&=& \frac{\gamma ^2}{16} \bigg(4 a_1^2-4 \sqrt{2} a_6 a_1+4 a_2^2+9 a_3^2+12 a_4^2+12 a_5^2+20 a_6^2\\
&&\hspace{20mm} +20 a_7^2+27 a_8^2-6 \sqrt{3} a_3-4 \sqrt{2} a_2 a_7-6 \sqrt{3} a_3 a_8-18 a_8\bigg) \notag
\end{eqnarray}
It is easy to check that this potential is minimized by exactly the steady state $\mathbf{a}_{\mathrm{SS}}$ given in Eq.~(\ref{dampSS}), as required.

\section{Conclusions}
\label{sec:Conclusions}

In this work, we have reformulated the finite-dimensional Lindblad equation as gradient flow. For Hermitian jump operators, the evolution admits a simple gradient flow structure with respect to a scalar potential $\Phi$, offering clear insight into the dynamics and steady-state behaviour. In the non-Hermitian case, the steady-state structure is still fixed by a potential $\Phi$, but the presence of a solenoidal term $R(\rho)$ complicates the dynamics. We demonstrated that one can still cast the equation in a gradient-like form via (i) a complexication of the potential or (ii) an orthogonal Helmholtz-Hodge decomposition. 
There is still work to be done in better understanding the more complicated dynamics of the non-Hermitian jump operator case. 

The algorithm described in Appendix \ref{App:Riccati} for constructing an orthogonal Helmholtz-Hodge decomposition is generic, making it well suited for numerical implementation, but must still be applied on a case-by-case basis. In practice, finding explicit expressions for the orthogonal choices of $\widetilde{\Phi}$ and $\widetilde{R}$ is difficult for generic jump operators. We leave the development of a general analytic method for this task to future work. 

Throughout this work, we have phrased everything in the interaction picture, absorbing the Hamiltonian into the definition of the jump operators. In realistic settings, interactions with an environment generically shift the free Hamiltonian, and incorporating these shifts into the interaction-picture jump operators is not always straightforward. As a result, the gradient flow structure may be modified by an additional von Neumann term, which can always be absorbed into the $R(\rho)$ contribution.

There are several other potentially interesting directions to explore:

\begin{itemize}

\item Thermal State Preparation: It would be valuable to understand whether thermal steady states can always be engineered by appropriately choosing the jump operators, and how special this feature is to qubits.
  
\item Numerical Applications: Since gradient-flow equations underlie many optimization methods (like gradient descent widely used in machine learning), they may provide useful intuition for numerical strategies. One could speculate that this structure might help in approximating or truncating the infinite hierarchy of equations that are often encountered in studies of adjoint master equations and mean-field approximations \cite{Plankensteiner:2021wst}. We leave this as an interesting direction for future work.
  
\item Alternative Metrics: We assumed a flat distance metric throughout in this work. Extending our framework to incorporate more physically natural metrics --- such as the Bures or Fubini-Study metrics --- could more clearly connect our approach to the entropic gradient flows studied in \cite{mittnenzweig2017entropic,carlen2017gradient,cao2019gradient,carlen2020non}.  The connection to the exact renormalization group (ERG) flows studied in \cite{Goldman:2024cvx} could also be interesting. There, the ERG was shown to be a Lindbladian quantum channel, and monotonicity of distinguishability followed from the data processing inequality. It would be interesting to relate this to our scalar potential $\Phi$ and its interpretation as a Lyapunov function.
  
\item Eigenstate Thermalization Hypothesis (ETH): In \cite{ODonovan:2024ocu}, Lindblad dynamics were derived from coupling to a finite-sized environment obeying ETH (see also \cite{Purkayastha:2024paj}). It would be interesting to explore whether our gradient flow formalism applies to such settings or to broader non-equilibrium problems related to ETH \cite{Moudgalya:2018rkm}. 
  
\item Non-Markovian extensions: We suspect that our approach can be generalized to non-Markovian systems. Non-Markovian master equations are notoriously difficult to solve, and in particular we suspect that the study of late-time states can be simplified by use of the formalism presented here. This approach could complement recent results such as those in \cite{DiMeglio:2024bir}.

\item Infinite-dimensional systems: Extending our formulation to infinite-dimensional systems involving quantum field theories would be very interesting as understanding steady states in these problems tends to be very challenging (for example in inflationary cosmology \cite{Colas:2022kfu,Colas:2024ysu,Colas:2024xjy}). It may be tractable to export the tools  presented here, as well as those involving Bloch vectors to this case using known results about  $\mathfrak{su}(\infty)$ \cite{Rankin:1991qr}.
  
\end{itemize}

We hope that our perspective on Lindblad evolution provides a useful framework for further analytical and numerical investigations across quantum information and open systems.

\section*{Acknowledgements}

We thank Patrick Hayden, Alejandro Kunold, and Arnab Pradhan for useful discussions. The authors are grateful to William Kretschmer and Shozab Qasim for comments on a preliminary version of the manuscript.


\begin{appendix}

\section{Some properties of $\wedge$ and $\star$} 
\label{App:WedgeStar}

In this appendix we collect some useful identities involving $\wedge$ and $\star$, some of which we have been unable to find in the literature. As emphasized in the main text, $\wedge$ and $\star$ are related to the operation of matrix commutation and anti-commutation respectively.
 
In the literature it seems like not much is written about the star product. Due to the complete symmetry of $d_{jk\ell}$ note that $\star$ is commutative such that $\mathbf{a} \star \mathbf{b} = \mathbf{b} \star \mathbf{a}$, however it is {\it not} associative $( \mathbf{a} \star \mathbf{b} ) \star \mathbf{c} \neq \mathbf{a} \star ( \mathbf{b} \star \mathbf{c} )$. It is distributive in the sense that $\mathbf{a} \star ( \mathbf{b} + \mathbf{c} ) = \mathbf{a} \star \mathbf{b} + \mathbf{a} \star \mathbf{c}$ and $(\mathbf{a} \star \mathbf{b} )  \cdot \mathbf{c} =  ( \mathbf{b} \star \mathbf{c} ) \cdot \mathbf{a}$. Finally from \cite{Bringewatt:2022suo} we get the non-trivial identity
\begin{equation}
(\mathbf{a} \star \mathbf{b}) \cdot (\mathbf{c} \star \mathbf{d}) =\big( (\mathbf{a} \star \mathbf{b} ) \star \mathbf{c} \big) \cdot \mathbf{d} \ .
\end{equation}
We also similarly define the wedge product $\wedge$ of two vectors $\mathbf{a}$ and $\mathbf{b}$ as
\begin{eqnarray} \label{wedge}
( \mathbf{a} \wedge \mathbf{b} )_{j} = f_{j k \ell} a_{k} b_{\ell} \ .
\end{eqnarray}
Notice that this is anti-commutative where $\mathbf{a} \wedge \mathbf{b}  =-  \mathbf{b} \wedge \mathbf{a}$ by complete antisymmetry of the $f_{jk\ell}$ (this property also implies that $\mathbf{b} \wedge \mathbf{b} =\mathbf{0}$). It is again {\it not} associative $( \mathbf{a} \wedge \mathbf{b} ) \wedge \mathbf{c} \neq \mathbf{a} \wedge ( \mathbf{b} \wedge \mathbf{c} )$ but is distributive where $\mathbf{a} \wedge ( \mathbf{b} + \mathbf{c} ) = \mathbf{a} \wedge \mathbf{b} + \mathbf{a} \wedge \mathbf{c}$ and $(\mathbf{a} \wedge \mathbf{b} ) \cdot \mathbf{c} =  ( \mathbf{b} \wedge \mathbf{c} ) \cdot \mathbf{a}$ (notice that the last identity implies $(\mathbf{a} \wedge \mathbf{b} ) \cdot \mathbf{b} = 0$ which means that the wedge product is a vector that lies {\it perpendicular} to the plane formed by $\mathbf{a}$ and $\mathbf{b}$). Finally we have the identity 
\begin{equation} \label{wedgewithdot}
(\mathbf{a} \wedge \mathbf{b}) \cdot (\mathbf{c} \wedge \mathbf{d}) = \big( (\mathbf{a} \wedge \mathbf{b} )  \wedge \mathbf{c} \big) \cdot \mathbf{d} \ .
\end{equation}
Notice that for $D=2$ and taking $f_{jk \ell} = \epsilon_{j k \ell}$ to be the Levi-Cevita symbol, then $\wedge$ becomes the standard cross product $\times$. If $d_{j k \ell} =0$ as is the case for the Pauli matrices in $D=2$ then there obviously is no $\star$ product.

There is also a family of Jacobi-like relations we can derive. The first for $\wedge$ follows from the usual commutator Jacobi relation
\begin{eqnarray}
\left[ X, [Y,Z] \right]+ \left[ Y, [Z,X] \right] + \left[ Z, [X,Y] \right] =0 \qquad \implies \qquad f_{ijm} f_{mkn} + f_{kim} f_{mjn} + f_{jkm} f_{min} = 0 
\end{eqnarray}
which when contracted with three vectors results in the identity
\begin{eqnarray}
( \mathbf{a} \wedge \mathbf{b} ) \wedge \mathbf{c} + ( \mathbf{c} \wedge \mathbf{a} ) \wedge \mathbf{b} + ( \mathbf{b} \wedge \mathbf{c} ) \wedge \mathbf{a} = \mathbf{0} \ .
\end{eqnarray}
There are however more related expressions one can find by stringing together commutators and anti-commutators of matrices as shown in \cite{Lavrov:2013mza}:
\begin{eqnarray}
\left[ X, \{Y,Z\} \right]+ \left[ Y, \{Z,X\} \right] + \left[ Z, \{X,Y\} \right] & = & 0\\
\left[ X, \{Y,Z\} \right] - \left\{ Z, [X,Y] \right\} + \left\{ Y, [ Z,X] \right\} & = & 0\\
\left[ X, [ Y, Z] \right]+ \left\{ Y, \{Z,X\} \right\} - \left\{ Z, \{X,Y\} \right\} & = & 0 \label{AAC}
\end{eqnarray}
This gives rise to the relationships:
\begin{eqnarray}
(\mathbf{a} \star \mathbf{b}) \wedge \mathbf{c} + (\mathbf{c} \star \mathbf{a}) \wedge \mathbf{b} + (\mathbf{b} \star \mathbf{c}) \wedge \mathbf{a} & = & 0 \label{Jac2} \\
(\mathbf{a} \wedge \mathbf{b}) \star \mathbf{c} - (\mathbf{c} \wedge \mathbf{a}) \star \mathbf{b} + (\mathbf{b} \star \mathbf{c}) \wedge \mathbf{a} & = & 0 \\
(\mathbf{a} \star \mathbf{b}) \star \mathbf{c} - (\mathbf{c} \star \mathbf{a}) \star \mathbf{b} - (\mathbf{b} \wedge \mathbf{c}) \wedge \mathbf{a} & = & \tfrac{2}{N} \big[ ( \mathbf{a} \cdot \mathbf{c} ) \mathbf{b} -  ( \mathbf{a} \cdot \mathbf{b} ) \mathbf{c} \big] \label{Jac4}
\end{eqnarray}

\section{Riccati and the orthogonal Helmoholtz-Hodge decomposition}
\label{App:Riccati}

The potential $\widetilde{\Phi}$ appearing in the orthogonal Helmholtz-Hodge decomposition can be represented in terms of Bloch vectors by Eq.~(\ref{mathcalPR}), repeated here,
\begin{eqnarray}
\widetilde{\Phi}(\mathbf{a}) := \sfrac{D}{2(D-1)} \mathbf{a}^{T} \mathbb{P} \mathbf{a} + \mathbf{p}^{T} \mathbf{a}
\end{eqnarray}
and one finds that the unknowns $\mathbb{P}$ and $\mathbf{p}$ satisfy the equations 
\begin{eqnarray}
\mathbb{M}^{T} \mathbb{P} + \mathbb{P}\mathbb{M} + 2 \mathbb{P}^2 & =& 0 \label{riccati1APP} \\
\mathbb{M}^{T} \mathbf{p} + \mathbb{P} ( \mathbf{v} + 2\mathbf{p}  )  & = & \mathbf{0} \label{riccati2APP} \\
\mathbf{p} \cdot (\mathbf{v} + \mathbf{p}) & = & 0 \label{riccati3APP}
\end{eqnarray}
with the additional constraint that
\begin{eqnarray}
\mathrm{Tr}[\mathbb{P}] = - \mathrm{Tr}[\mathbb{M}] \ .
\end{eqnarray}
The equation (\ref{riccati1APP}) turns out to be an algebraic Ricatti equation which is degenerate (since the equation is homogeneous). These equations are difficult to solve generically, which can be seen by the fact that Eq.~(\ref{riccati1APP}) has at most $2^{D^2 - 1}$ solutions.

There turns out to be a generic procedure for solving the degenerate case, as described in \cite{kesavan2009degenerate}, however for the particular problem at hand this procedure turns out to give the trivial solution $\mathbb{P} = 0 $ (an instance of what is known as a ``stabilizing solution'' in the optimal control theory literature, for reasons unrelated to behavior of the steady states of the Lindblad equation).

It turns out that there is a simple solution to this problem, which only slightly tweaks the usual control theory procedure for finding stabilizing solutions \cite{laub2003schur,byers1987solving}. To find the desired solution of Eq.~(\ref{riccati1APP}) one considers the enlarged $(2D^2-2) \times (2 D^2-2) $ matrix, in block matrix form,
\begin{equation} 
\mathbb{Z} = \left[ \begin{matrix} \mathbb{M} & - 2 \mathbb{I}_{D^2 -1} \\ 0 & - \mathbb{M}^{T} \end{matrix} \right] \ .
\end{equation}
One must then perform a Schur decomposition of the matrix $\mathbb{Z}$, which puts it in the form
\begin{eqnarray}
\mathbb{Z} = \mathbb{U} \mathbb{T} \mathbb{U}^{T}
\end{eqnarray}
where $\mathbb{T}$ is an upper triangular matrix and $\mathbb{U}$ is an orthogonal matrix (satisfying $\mathbb{U}^{T} \mathbb{U} = \mathbb{I}_{2D^2 -2}$), with the block structure
\begin{equation} 
\mathbb{U} = \left[ \begin{matrix} \mathbb{U}_{11} & \mathbb{U}_{12} \\  \mathbb{U}_{21}  & \mathbb{U}_{22} \end{matrix} \right] \qquad \mathrm{and} \qquad \mathbb{T} =  \left[ \begin{matrix} \mathbb{T}_{11} & \mathbb{T}_{12} \\  0  & \mathbb{T}_{22} \end{matrix} \right]  \ .
\end{equation}
Because $\mathbb{M}$ comes from a Lindblad equation, we use without loss of generality that the matrix $\mathbb{M}$ has $D^2 - 1$ non-positive eigenvalues $\lambda_j$, which means that the matrix $\mathbb{T}$ has $D^2 - 1$ pairs of real eigenvalues $\pm \lambda_j$. In order to find the solution, one orders\footnote{Those familiar with the algorithm of finding usual stabilizing solutions \cite{laub2003schur,byers1987solving} of the Riccati equation will recognize that we perform this ordering in the {\it opposite} direction.} the eigenvalues of $\mathbb{T}$ so that the positive eigenvalues are contained in $\mathbb{T}_{11}$ and the negative eigenvalues are contained in $\mathbb{T}_{22}$, and where one orders the eigenvalues in decreasing order throughout. It then turns out that the desired solution is given by 
\begin{equation} 
\mathbb{P} = \mathbb{U}_{21} \mathbb{U}_{11}^{-1} \ .
\end{equation}
One may then check that $\mathrm{Tr}[\mathbb{P}] = -\mathrm{Tr}[\mathbb{M}]$ for this particular solution. Once this matrix is known, it is then trivial to solve for the vector $\mathbf{p}$ using Eq.~(\ref{riccati2APP}), and the solution automatically satisfies Eq.~(\ref{riccati3APP}).

We have confirmed that this procedure performs reliably in numerical implementations, although we do not provide a formal proof of its validity. We expect it may be of interest to researchers pursuing algorithmic approaches to orthogonal Helmholtz-Hodge decompositions, as emphasized in \cite{suda2020application}, where the need for a generic construction was explicitly noted.

\end{appendix}

\bibliography{references}

\end{document}